\documentclass[11pt,showpacs,showkeys]{revtex4}
\usepackage{graphicx}

\begin{document}

\title{Breakdown of Migdal--Eliashberg theory via catastrophic vertex
divergence at low phonon frequency}

\author{J.P.Hague}
\affiliation{Department of Physics, University of Warwick, CV4 7AL, U.K.}
\affiliation{Department of Physics, Loughborough University, LE11 3TU, U.K.}
\email{J.P.Hague@lboro.ac.uk}

\author{N.d'Ambrumenil}
\affiliation{Department of Physics, University of Warwick, CV4 7AL, U.K.}

\date{25/09/2007}

\begin{abstract}
We investigate the applicability of Migdal--Eliashberg (ME) theory by
revisiting Migdal's analysis within the dynamical mean-field theory
framework. First, we compute spectral functions, the quasi-particle
weight, the self energy, renormalised phonon frequency and resistivity
curves of the half-filled Holstein model. We demonstrate how ME theory
has a phase-transition-like instability at intermediate coupling, and
how the Engelsberg--Schrieffer (ES) picture is complicated by
low-energy excitations from higher order diagrams (demonstrating that
ES theory is a very weak coupling approach). Through consideration of
the lowest-order vertex correction, we analyse the applicability of ME
theory close to this transition. We find a breakdown of the
theory in the intermediate coupling adiabatic limit due to a
divergence in the vertex function. The region of applicability is
mapped out, and it is found that ME theory is only reliable in the
weak coupling adiabatic limit, raising questions about the accuracy of
recent analyses of cuprate superconductors which do not include vertex
corrections. 
\end{abstract}

\maketitle

\section{Introduction}

\label{sct:intro}

The use of Migdal--Eliashberg (ME) theory for the study and analysis
of electron-phonon systems is widespread. For conventional
superconductors such as lead, where the electron-phonon coupling is
higher than can be treated with BCS theory, ME theory has been
extremely successful for understanding the superconducting
properties. Recently, a lot of researchers have been interested in an
apparent kink in the electronic dispersion of cuprate supercondutors
as determined from angle resolved photo-emission spectroscopy
(ARPES), and their analysis / interpretation typically uses the
related Engelsberg--Schrieffer (ES) result (i.e. just the lowest order
Fock diagram) \cite{engelsberg, lanzara1,cuk2004a}. The ES result
divides excitations into long-lived (coherent) low-energy excitations,
and rapidly decaying high-energy excitations, with a kink at the
phonon energy. ES based analysis of ARPES results suggests a large
electron-phonon coupling in the cuprates, with estimates of the
dimensionless coupling constant lying between $\sim 0.3$ and $1.5$
(depending on doping) and very large phonon frequencies of $\sim 50$
meV \cite{lanzara1}. There has also been a development of a maximum
entropy technique for analysing cuprate superconductors which makes
use of Eliashberg theory \cite{hwang1}. The analysis in
Ref. \onlinecite{hwang1} determines a dimensionless electron-phonon
coupling of $\lambda>1$, but this is well above the $\lambda\sim 1$
value where one would normally expect perturbation theory to fail. The
region of intermediate electron-phonon coupling is also relevant to a
number of other materials. For example, electron paramagnetic
resonance measurements on the manganites support a strong
``electron-phonon'' coupling leading to Jahn--Teller polarons
\cite{zhao1}. In light of the current experimental situation, and the
importance of the conclusions of Refs. \onlinecite{cuk2004a, hwang1},
a full re-analysis of the perturbation theory is of high significance.

The neglect of vertex corrections suggested by Migdal leads to a
theory where an infinite set of Feynman diagrams may be
summed. According to an analysis carried out by Migdal, this theory
should be valid in the physical regime of electron-phonon problems,
where the phonon energy is significantly smaller than the intersite
hopping (``Migdal's theorem'') \cite{migdal1, eliashberg1}
specifically that the vertex corrections are small when
$\lambda\omega_0/\epsilon_f \ll 1$ ($\omega_0$ is the phonon frequency
and $\epsilon_f$ the Fermi energy)
\footnote{Migdal's theorem it is not a theorem in the mathematical
sense, so we prefer to call it Migdal's analysis in this article.}. On
the basis of Migdal's analysis, it is often believed that
Migdal--Eliashberg theory is applicable above $\lambda\sim 1$ in the
adiabatic limit because of the apparently small size of the vertex
corrections, even though in general perturbative approaches break down
(i.e. the functional form of the self-energy becomes incorrect) when
the coupling constant becomes large. In this article, we revisit
Migdal's analysis by examining the large-$d$ limit (local
approximation).

The large-$d$ limit has been an effective workshop for determining the
validity of approximate schemes. The quantum Monte-Carlo (QMC)
solution in the large-$d$ limit \cite{freericks14,hirsch2} has been
compared with a number of different diagrammatic approaches
\cite{freericks1,freericks2,freericks3} demonstrating that
self-consistent second-order perturbation theory worked better than
Migdal-Eliashberg theory at intermediate phonon frequencies and strong
coupling ($\omega _{0}\sim 0.2t$, $ U>t $). The QMC solution is
difficult for very low phonon frequency and low temperature. QMC
self-energies and Green functions are generated along the Matsubara
axis. As such, it is not easy to make quantitative conclusions about
the point of breakdown of ME theory as small differences at individual
Matsubara frequencies may result in relatively large differences in
the spectral functions. Using an alternative approach, Benedetti
{\it et al.} \cite{benedetti1} found the formation of more than one
extremum in the path-integral formulation at very low phonon
frequencies and intermediate coupling, leading to a breakdown of ME
theory. Work on the finite-dimensional Holstein model also shows
deficiencies in ME theory. On the basis of a comparison with exact
diagonalization results, Alexandrov {\it et al.} \cite{alexandrov1}
have shown that ME theory may break down at intermediate couplings
even in the adiabatic limit.

In this article, we investigate the Holstein model of electron-phonon
interactions \cite{holstein1}, which employs a number of
approximations. The phonon dispersion is flat corresponding to
independently moving ions and phonon anharmonicity is neglected,
resulting in a Hamiltonian,
\begin{equation}
\label{eqn:holsteinhamiltonian}
H=-t\sum _{<ij>\sigma }c^{\dagger }_{i\sigma }c_{j\sigma }+\sum _{i\sigma }(gx_{i}-\mu )n_{i\sigma }+\frac{M\omega _{0}^{2}}{2}\sum _{i}x_{i}^{2}+\frac{1}{2M}\sum _{i}p_{i}^{2},
\end{equation}
 where $c^{\dagger }_{i\sigma }$ $( c_{i\sigma} )$ create
(annihilate) electrons at site $i$ with spin $\sigma$, $x_{i}$ is the
local ion displacement, $p_{i}$ the ion momentum, $M$ the ion mass,
$t$ the electron hopping parameter and $\mu$ the chemical potential. A
mean-field Coulomb pseudopotential may also be included, but in the
normal state, it is just absorbed into the chemical potential. In the
following, $t=0.5$ and all energies (including temperature) are
measured in units of $2t$.

An expression for the effective interaction between electrons can be
obtained by performing a linear transformation in the phonon variable
to remove the electron-phonon term \cite{bickers1}. The resulting
interaction is retarded, with Fourier components:
\begin{equation}
\label{eqn:barephonprop}
U(i\omega _{s})=-\left( \frac{g^{2}}{M\omega _{0}^{2}}\right)
\frac{\omega _{0}^{2}}{(\omega _{s}^{2}+\omega _{0}^{2})} = -U
\frac{\omega _{0}^{2}}{(\omega _{s}^{2}+\omega _{0}^{2})},
\end{equation}
where $\omega _{s}=2\pi sT$ are the Matsubara frequencies for
Bosons. Taking the limit $\omega _{0}\rightarrow \infty$,
$g\rightarrow \infty$, while keeping the ratio $g/\omega _{0}$ finite,
leads to an attractive Hubbard model \cite{hubbard1} with
instantaneous attraction of magnitude $U = g^{2}/M\omega_{0}^{2}\equiv
W\lambda$, where $W$ is the half band-width (note that $U$ is related
to the bipolaron binding energy for the strong coupling two-electron
problem). In the opposite limit ($\omega _{0}\rightarrow 0$,
$M\rightarrow \infty$, keeping $M\omega _{0}^{2}\equiv \kappa$ finite)
the phonon kinetic-energy vanishes, leaving only a static variable
$x_{i}$ representing the phonon subsystem in the Hamiltonian, $
H=-t\sum _{<ij>\sigma }c^{\dagger }_{i\sigma }c_{j\sigma }+\sum
_{i\sigma }(gx_{i}-\mu )n_{i\sigma }+\frac{\kappa }{2}\sum
_{i}x_{i}^{2}$.  Thus the problem looks like that of a single electron
in a disordered potential. The large-$d$ limit of this model was
exactly solved by Millis {\it et al.} \cite{millis2} and extended to
deal with long range order by Ciuchi {\it et al} \cite{ciuchi2}. Thus,
the phonon frequency may be thought of as a parameter for tuning the
level of electronic correlation \cite{haguetuning}.

In this paper, we investigate the validity of the Migdal--Eliashberg
approach within the dynamical mean-field theory formalism. As we carry
out the self-consistency on the real-axis self-energies, we can
investigate the behaviour of ME theory over a full range of
temperatures and phonon frequencies. We begin by calculating
self-energies, spectral functions, the quasi-particle weight and
renormalized phonon frequency.  To determine the validity of the
theory, we calculate an expression for the lowest-order correction to
the vertex function, and evaluate its magnitude.  We find that the
theory can also break down in the low frequency regime, contrary to
the standard interpretation of Migdal's analysis. Finally, we
calculate resistivity curves in the regime where the lowest-order
corrections are small.

\section{Migdal--Eliashberg theory in the local approximation}

\label{sct:dmft}

In this article, we use the local approximation or dynamical
mean-field theory (DMFT) as a way of analysing the Migdal-Eliashberg
theory. The self-energy of correlated-electron systems is momentum
independent in limit of large dimensions \cite{metzner1}, and
approximately momentum independent in 3D. In large-$d$, lattice models
map onto Anderson impurity models with a self-consistent hybridisation
\cite{georges4}.

The self-consistent DMFT equations may be obtained by rewriting the
action for the model to be considered in terms of an effective
single-site action \cite{georges1}
\begin{equation}
\label{eqn:effectiveaction}
S_{\mathrm{eff}}=-\int _{0}^{\beta }d\tau \int _{0}^{\beta }d\tau '\sum _{\sigma }c_{i\sigma }^{\dagger }(\tau ){\mathcal{G}}_{0}^{-1}(\tau -\tau ')c_{i\sigma }(\tau ')+S_{\mathrm{int}}.
\end{equation}
Here ${\mathcal{G}}_{0}(i\omega _{n})$ plays the r\^{o}le of the
host Green function in the equivalent impurity model. If one assumes
that correlations carried via the bath between electrons entering and
exiting a single site can be neglected (true in the case of large
coordination number), the degrees of freedom associated with all but
one site can be integrated out. This leads to the following
self-consistent equation \cite{georges1},
\begin{equation}
\label{eqn:selfconsistent}
{\mathcal{G}}_{0}^{-1}(i\omega _{n})=i\omega _{n}+\mu +G^{-1}(i\omega _{n})-{R[G(i\omega _{n})]},
\end{equation}
 where $G(i\omega _{n})$ is the site-local (impurity) Green
function, and is itself a functional of ${\mathcal{G}}_{0}(i\omega
_{n})$. $\omega _{n}=2\pi T(n+1/2)$ are the Fermionic Matsubara
frequencies. $R[x]$ is the reciprocal function of the Hilbert
transform, defined as $R[\tilde{D}(\xi )]=\xi$.  The Hilbert transform
of the non-interacting density of states, \( {\mathcal{D}}(\varepsilon
) \), is defined as $\tilde{D}(\xi )\equiv \int _{-\infty }^{+\infty
}d\varepsilon {\mathcal{D}}(\varepsilon )/(\xi -\varepsilon)$.

When electrons move on a tight-binding hypercubic lattice, the bare
DOS takes the form of a Gaussian \cite{metzner1},
${\mathcal{D}}(\varepsilon )=\exp (-\varepsilon
^{2}/2t^{2})/t\sqrt{2\pi }$ and there is no simple expression for
the reciprocal function. Introducing the modified Dyson equation,
\begin{equation}
\label{eqn:weissfield}
{\mathcal{G}}_{0}^{-1}(i\omega _{n})=G^{-1}(i\omega _{n})+\Sigma (i\omega _{n}),
\end{equation}
 (where $\Sigma (i\omega _{n})$ is the electron self-energy), equation
(\ref{eqn:selfconsistent}) may be rewritten as, 
\begin{equation}
R[G(i\omega _{n})]=i\omega _{n}+\mu -\Sigma (i\omega _{n}),
\end{equation}
 and inverted to give an expression for the Green function in terms
of the self-energy, 
\begin{equation}
\label{eqn:greensfn}
G(i\omega _{n})=\int \frac{d\varepsilon \, {\mathcal{D}}(\varepsilon )}{i\omega _{n}+\mu -\Sigma (i\omega _{n})-\varepsilon }.
\end{equation}
This also allows the approximation to be interpreted as a course
graining of the momentum space \cite{hettler1}. To complete the
scheme, a form for the electronic self-energy must be calculated in
terms of the non-interacting Weiss field. This is normally
approximate. Then a self-consistent procedure is followed: Compute the
Green function from equation (\ref{eqn:greensfn}), the Weiss field
from equation (\ref{eqn:weissfield}) and then re-calculate the
self-energy until convergence is reached.

\begin{figure}
\includegraphics[width=13cm]{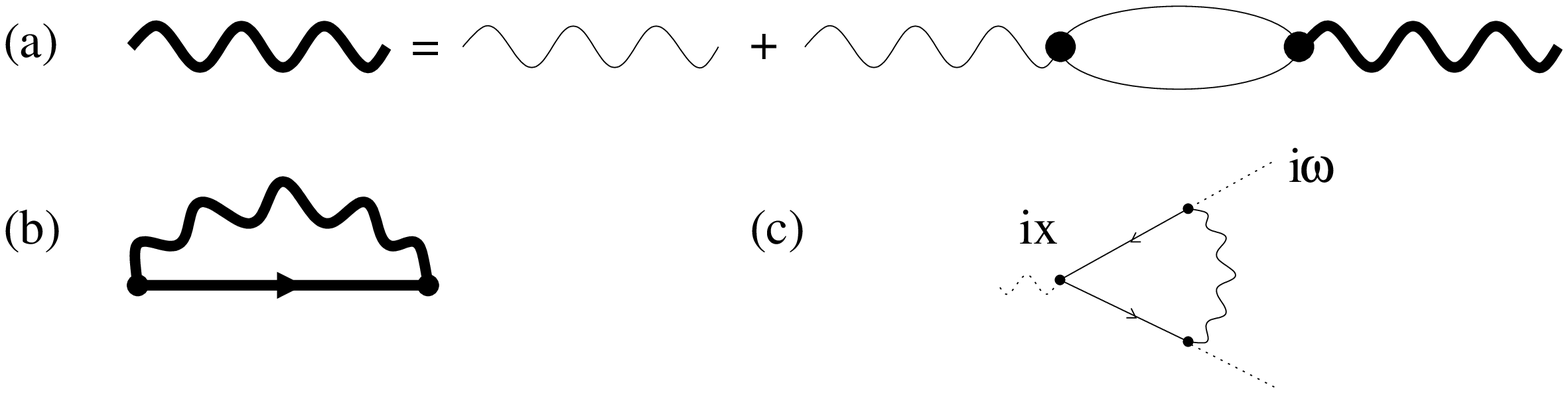}  
\caption{(a) Dyson's equation for the phonon Green function to second
order when vertex corrections are neglected. (b) Renormalized Fock
diagram. (c) The first-order correction to the vertex
function. \protect\( ix\protect \) is the frequency of the emitted
phonon and \protect\( i\omega \protect \) is the frequency of the
incoming electron. Neglect of this diagram is central to Migdal's
theorem.  Thick and thin lines represent the full and bare Green
functions respectively.  Wavy lines represent phonons and straight
lines electrons.\label{fig:feynmandiag}}
\end{figure}

The application of ME theory within DMFT corresponds to computing the
self-energy and phonon propagator from the diagrammatic equations in
figure \ref{fig:feynmandiag}(a and b). The self-consistent solution of
such a self-energy corresponds to the summation of all Feynman
diagrams which contain no vertex corrections. In the low (non-zero)
phonon-frequency limit, Migdal's analysis indicates a condition, \(
U\omega _{0}\ll t^{2} \), for the neglect of corrections to the vertex
function \cite{migdal1}.

The solution of the Dyson equation seen in figure \ref{fig:feynmandiag}(a)
results in the full phonon propagator, 
\begin{equation}
\label{eqn:dysonphon}
D(\omega )=\frac{\omega _{0}^{2}}{(\omega +i\eta )^{2}-\omega
_{0}^{2}[1+\Pi _{0}(\omega )]},
\end{equation}
 which is then used for the calculation of the electron
self-energy. The phonon polarisation bubble (dimensionless
self-energy) $\Pi _{0}(i\omega _{s})$ is given from perturbation
theory,
\begin{equation}
\label{eqn:matsubarapiphon}
\Pi _{0}(i\omega _{s})=-2UT\sum _{m}G(i\omega _{m})G(i\omega
_{s}+i\omega _{m}),
\end{equation}
 and may be analytically continued by introducing the spectral
representation, $G(i\omega _{n})=\int dx\rho (x)/(i\omega _{n}-x)$
where $\rho (\omega )=\mathrm{Im}[G(\omega +i\eta )]/\pi$
and performing the sum over Matsubara frequencies to give,
\begin{equation}
\label{eqn:pizero}
\mathrm{Im}[\Pi _{0}(\omega )]=-2U\int _{0}^{\omega }dx\rho (-x)\rho (\omega -x),
\end{equation}
 at absolute zero, and
\begin{equation}
\label{eqn:pitemp}
\mathrm{Im}[\Pi _{0}(\omega,T )]=-U\sinh (\frac{\omega }{2T})\int _{-\infty }^{\infty }dx\, \frac{\rho (-x)}{\cosh (\frac{x}{2T})}\frac{\rho (\omega -x)}{\cosh (\frac{\omega -x}{2T})},
\end{equation}
at finite $T$. The full spectral function is used so that all diagrams
with no vertex corrections are included, consistent with the proper interpretation of Migdal's analysis.

The lowest-order skeleton diagram shown in figure
\ref{fig:feynmandiag}(b),
\begin{equation}
\Sigma (i\omega _{n})=-UT\sum _{s}G(i\omega _{n}-i\omega _{s})D(i\omega _{s}),
\end{equation}
may be analytically continued in the same way as the polarisation bubble
to give, 
\begin{equation}
\label{eqn:selfenergyeqn}
\mathrm{Im}[\Sigma (\omega )]=U\int _{0}^{\omega }dx\rho (x)\sigma
(\omega -x),
\end{equation}
 with $\sigma (\omega )=\mathrm{Im}[D(\omega +i\eta )]/\pi$. A more complicated
expression applies at finite temperature,
\begin{equation}
\label{eqn:selfenergyt}
\mathrm{Im}[\Sigma (\omega )] = U\cosh (\frac{\omega }{2T})\int _{-\infty }^{\infty }\frac{dx\, \rho (\omega -x)\sigma (x)}{2\cosh (\frac{\omega -x}{2T})\sinh (\frac{x}{2T})}.
\end{equation}
 At each stage, the Kramers--Kronig relation is used to compute the
real parts of the electron and phonon self-energies, for instance
$\mathrm{Re}[\Pi _{0}(\omega )]=\mathrm{P}\int
dx \mathrm{Im}[\Pi _{0}(x)]/\pi(\omega -x)$, where P denotes the
principal integral.

After the self-consistent procedure has converged, physical properties
are calculated, including the quasi-particle weight (inverse effective
mass),
\begin{equation}
\label{eqn:effmass}
Z=\frac{m_{0}}{m^{*}}=1-\left. \frac{\partial \Sigma }{\partial \omega }\right| _{\omega =0},
\end{equation}
 the effective phonon frequency $\Omega$ from, 
\begin{equation}
\label{eqn:phonfreq}
\Omega ^{2}-\omega _{0}^{2}(1+\mathrm{Re}[\Pi _{0}(\Omega )])=0,
\end{equation}
 and the optical conductivity \cite{pruschke1}, 
\begin{equation}
\label{eqn:conductivity}
\mathrm{Re}[\sigma (\omega )]=\frac{\pi }{\omega }\int _{-\infty }^{\infty }d\varepsilon {\mathcal{D}}(\varepsilon )\int _{-\infty }^{\infty }d\nu \, \rho (\varepsilon ,\nu )\rho (\varepsilon ,\nu +\omega )[f(\nu )-f(\nu +\omega )],
\end{equation}
 where $f(x)$ is the Fermi-Dirac distribution and $\rho(\epsilon,\nu)
= \mathrm{Im}[1/(\nu+i\eta-\epsilon-\Sigma(\nu))]/\pi$ (taking the
limit, $\omega \rightarrow d\omega$, the DC conductivity is
recovered.)

Although DMFT is approximate, the formalism can be expected to work in
the non-interacting limit, where course graining will give the exact
non-interacting DOS for the tight-binding model (regardless of
dimension). In the opposite limit ($t\rightarrow 0$), the neglect
of loops through the host is justified, and the formalism should be
exact. Such propagation of correlation is also small when the
coordination number is high (e.g. FCC lattices).

\section{Spectral properties}

\label{sct:metheory}

\begin{figure}
\includegraphics[height=10cm,angle=270]{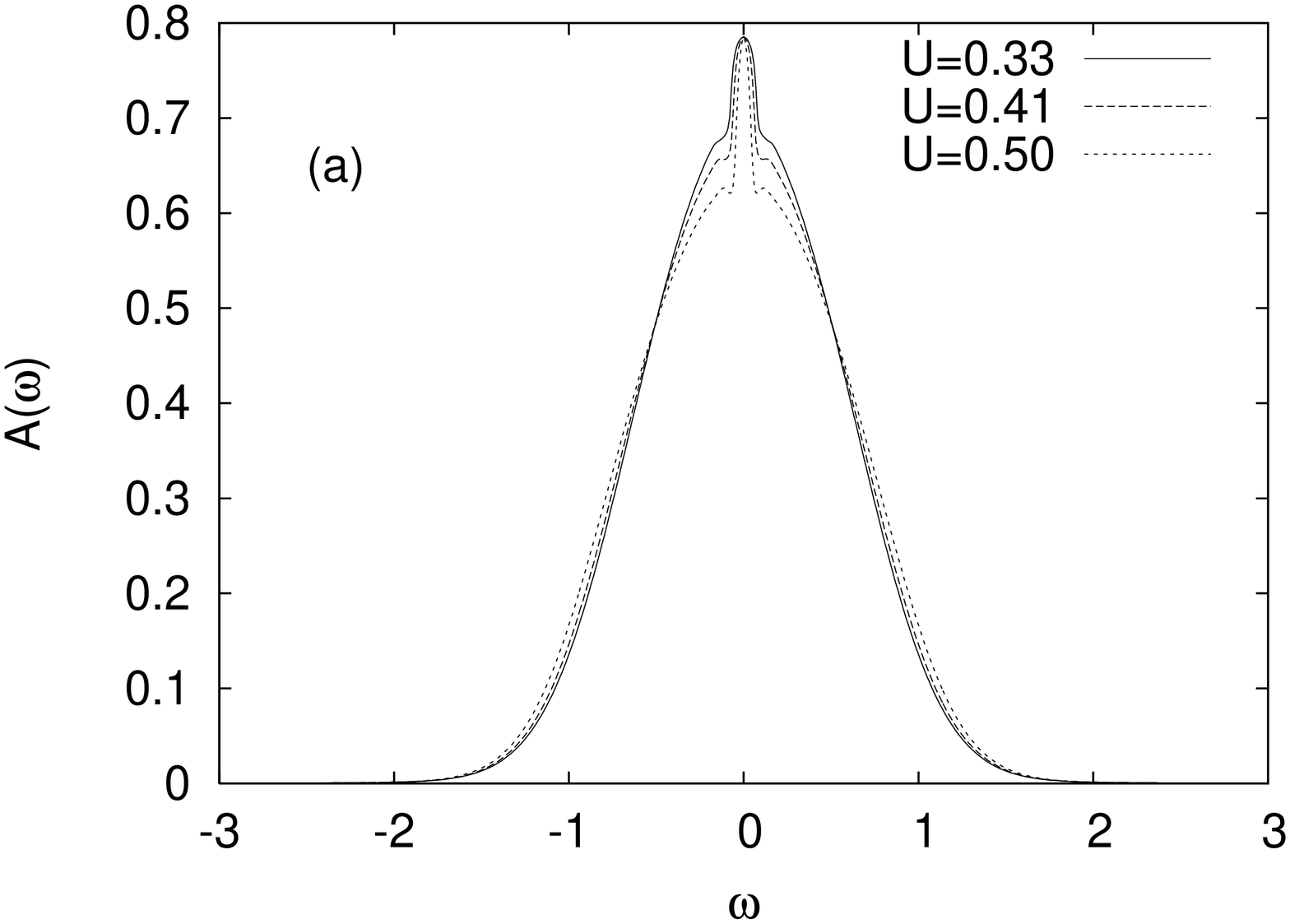}  
\includegraphics[height=10cm,angle=270]{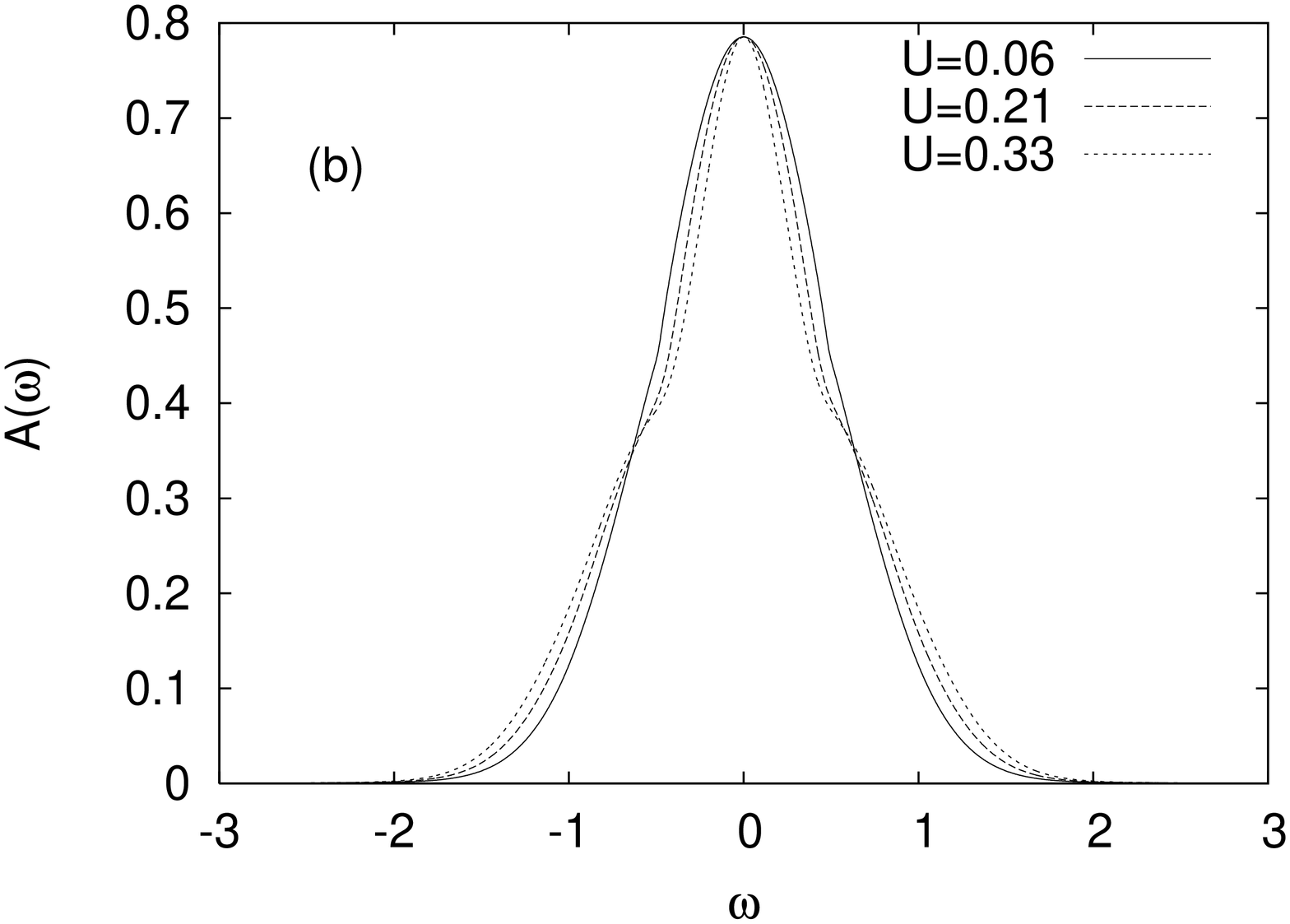}  
\caption{Spectral functions of the Holstein model computed using
Migdal-Eliashberg theory for $\omega _{0}=0.125$ (panel a) and
$\omega_{0}=0.5$ (panel b) at $T=0$.  A bi-polaronic resonance forms
at zero frequency. No upper and lower sub-bands are formed but
spectral weight is shifted away from the Fermi-energy.  The central
peak narrows with increasing $U$, corresponding to a divergence in the
effective mass. The general form of the curves is similar at all
frequencies.\label{fig:spectralfunctions}}
\end{figure}

We have solved the DMFT equations using the self-energies in equations
(\ref{eqn:selfenergyeqn}) and (\ref{eqn:selfenergyt}) to find electron
spectral functions. We show the evolution of these functions with
coupling for $\omega _{0} = 0.125$ in figure
\ref{fig:spectralfunctions}(a) and $\omega_0=0.5$ in figure
\ref{fig:spectralfunctions}(b). The spectral functions show two
features. There is a low energy peak with a width defined by the
phonon frequency, and a high energy shoulder. This behaviour can be
related to the two limits of the Holstein model. At low energy scales
(\( \omega <\omega _{0} \)), electrons interact via virtual processes
and the behaviour is essentially Hubbard-like (correlated). At
frequencies greater than \( \omega _{0} \), the spectral weight is
reduced, phonons may be created, and static behaviour emerges (in the
static limit, phonons may always be created as no energy is required
to deform the lattice). The central peak narrows with increased
coupling, until a critical value is reached. Here, the theory has a
Brinkman--Rice-like transition with a diverging effective mass
\cite{brinkman1}.

\begin{figure}
\includegraphics[height=10cm,angle=270]{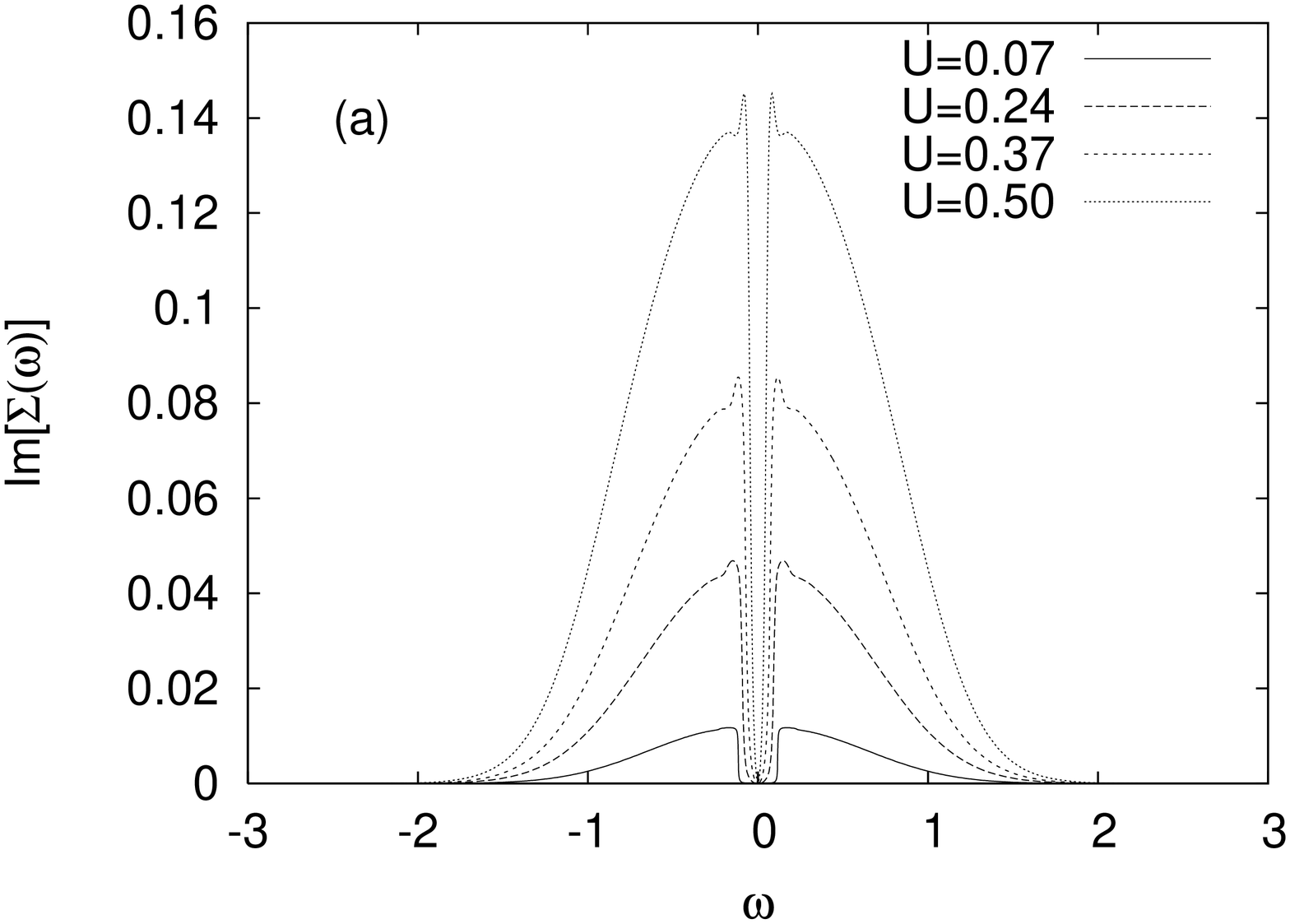} 
\includegraphics[height=10cm,angle=270]{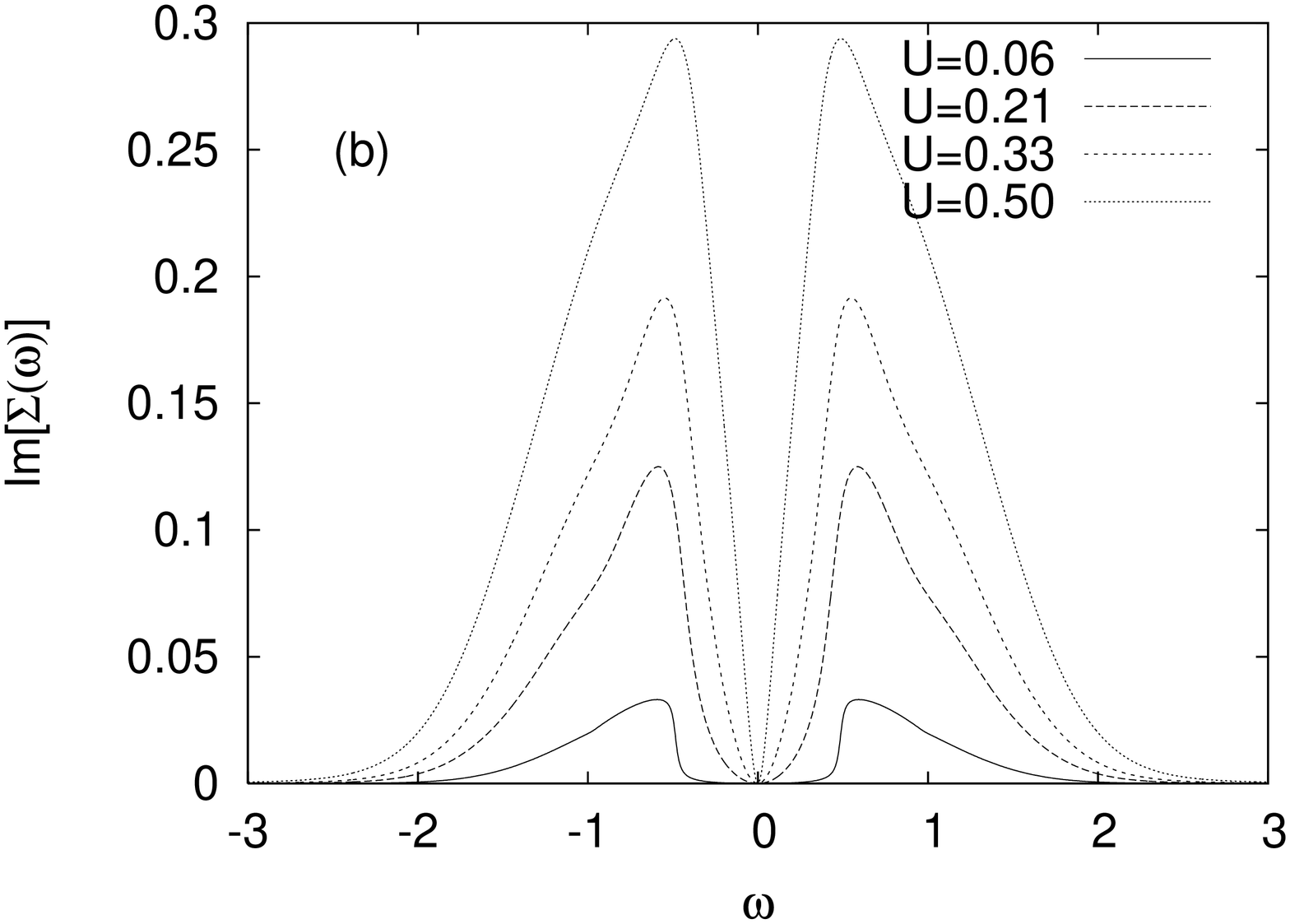} 
\caption{Self-energy of the Holstein model solved using ME theory for
$\omega _{0}=0.125$ (panel a) and $\omega_{0}=0.5$ (panel b) at $T=0$. As the
coupling, $U = W\lambda$ (see discussion after eqn 2) increases,
Hubbard-like behaviour is seen at low energy scales, and FK-like
behaviour at \protect\( |\omega |>\omega _{0}=0.2\protect \). Note how
the gap in the self-energy at weak coupling (which is central to the
Engelsberg--Schrieffer approach) fills up as coupling is
increased. This is due to the increasing importance of higher order
diagrams at stronger couplings. This shows that one should be careful
about interpreting results from the cuprates ($\lambda\gtrsim 1$)
using an ES approach (this type of analysis is typically used on ARPES
results), since the ES form for the self-energy (i.e. gapped at
$|\omega|<\omega_0$) can only be seen for very small
$\lambda$. \label{fig:selfenergy}}
\end{figure}

In figure \ref{fig:selfenergy}, we plot the imaginary part of the
self-energy.  For low coupling strengths (small $U$ and energy scales
($|\omega |<\omega _{0}$), \( \mathrm{Im}[\Sigma (\omega )] \) is
small because electrons cannot create phonons, consistent with the
Englesberg--Schrieffer analysis). However, as the coupling increases,
electrons can be scattered by bipolaron resonances and \(
\mathrm{Im}[\Sigma (\omega )] \) rises sharply on either side of the
Fermi-energy. The gap in the weight of the self-energy is frequently
assumed in the analysis of ARPES data, but as we can see here, when
the electron-phonon coupling approaches the band width
(i.e. $\lambda\sim 1$), the self-energy has a more complicated form,
without the simple picture of coherent and incoherent
quasi-particles. Indeed, it has been demonstrated that in low
dimensions, vertex corrections are required to achieve a sharp
discrimination between coherent and incoherent dressed electrons
\cite{hague2003a}.

\begin{figure}
\includegraphics[height=10cm,angle=270]{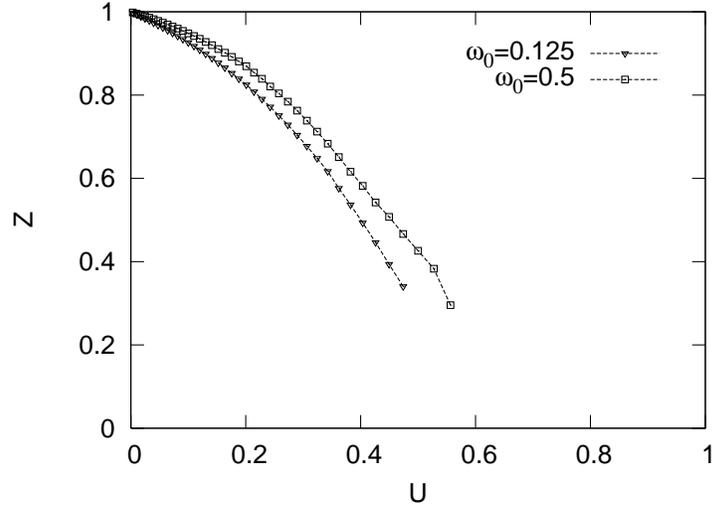} 
\caption{Quasi-particle weight, $Z$, computed using ME theory as a
function of $U$ (results for phonon frequencies $\omega_0 = 0.125$ and
$\omega_0 = 0.5$ are shown). $Z$ is strongly renormalised for
intermediate $U$.\label{fig:qpweight}}
\end{figure}

The onset of this ``transition'' can be studied by examining the inverse
quasi-particle mass (quasi-particle weight), shown in figure
\ref{fig:qpweight} for two values of phonon frequency. As the coupling
increases, \( Z \) becomes smaller and eventually vanishes at a
critical value of coupling (the effective mass diverges).

\begin{figure}
\includegraphics[height=10cm, angle=270]{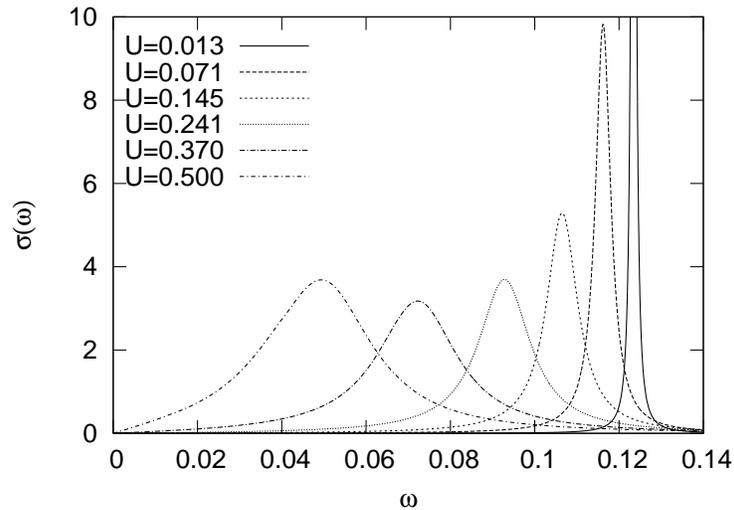}
\caption{Dependence of the phonon spectral function on coupling at
$\omega _{0}=0.125$. The effective phonon frequency (location of
maximum) and particle lifetime (inverse width of peak) are reduced
with increasing coupling. When $U=0.5$ the curve is skewed in such a
way that it is not Lorenzian, and the excitation no longer has the
properties of a single phonon.\label{fig:phonspec}}
\end{figure}

\begin{figure}
\includegraphics[height=10cm, angle=270]{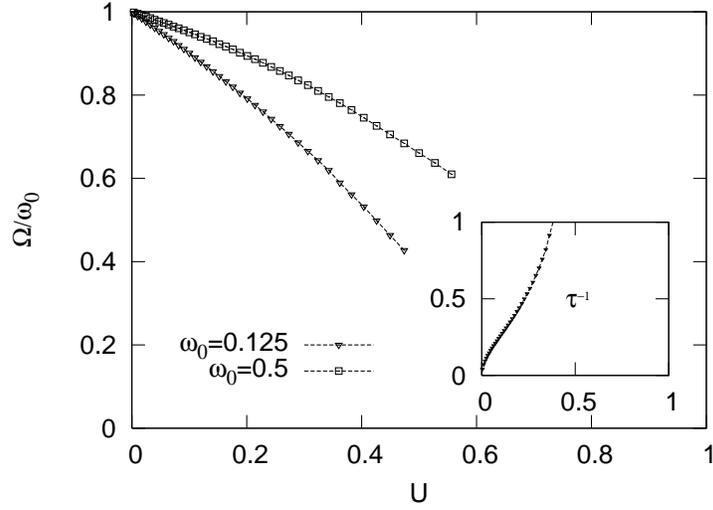}
\caption{Coupling dependence of the renormalised phonon frequency
defined by $\Omega ^{2}-\omega _{0}^{2}(1+\mathrm{Re}[\Pi _{0}(\Omega
)])=0$.  The renormalisation of the effective phonon frequency is not
as strong as that of the electronic inverse mass (quasi-particle
weight $Z$) shown in figure \ref{fig:qpweight}, which goes to zero at
the transition. The inset shows the quasiparticle lifetime given by
\protect\( \sqrt{\mathrm{Im}[\Pi _{0}(\Omega )]}\protect \). When the
value of $\lambda$ tends to 1, the phonons can no longer be treated as
single particle excitations. \label{fig:phonfreq}}
\end{figure}

We also examine the phonon spectral function in figure
\ref{fig:phonspec} for $\omega _{0}=0.125$ and various couplings. As
the coupling is increased, the phonon modes soften (figure
\ref{fig:phonfreq}). The effective frequency does not tend to zero as
quickly as the quasi-particle weight. We will revisit this point later
in this article.

It is clearly the case that ME theory does not correctly describe the
strong coupling limit. It is known from the exact solution of the
static limit \cite{millis2} and from approximate
``iterated-perturbation theory'' \cite{georges2} and QMC
\cite{jarrell1} solutions of the Hubbard model, that sub-bands should
form at strong coupling.  As our calculations demonstrate, this
sub-band formation is not properly reproduced within ME theory and
this makes it likely that effects of vertex corrections become
significantly more important at strong coupling. Since this is at
variance with the traditional interpretation of Migdal's analysis, we
are motivated to re-examine the vertex function in the next section.

\section{Breakdown of Migdal--Eliashberg theory}

\label{sct:break}

The neglect of the lowest-order vertex correction shown in figure
\ref{fig:feynmandiag}(c) and all higher-order corrections is central
to Migdal--Eliashberg theory. In this section, we compute the
lowest-order correction and use this to define the region of validity
for ME theory. In this sense, we are revisiting Migdal's analysis to
understand why there is a contradiction between low frequency results
computed with ME theory and advanced numerical methods.

The ratio of first to zeroth order vertices at finite temperature may
be written as 
\begin{equation} 
\frac{\Gamma _{1}(i\omega _{n},i\omega
_{x})}{\Gamma _{0}}=TU\sum _{s}{\mathcal{G}}_{0}(i\omega _{n}-i\omega
_{s}){\mathcal{G}}_{0}(i\omega _{n}-i\omega _{s}-i\omega
_{x})D_{0}(i\omega _{s}).
\end{equation}

As temperature tends to zero, this sum becomes an integral which may
easily be evaluated ($x$ and $\omega$ are defined as in figure
\ref{fig:feynmandiag}),
\begin{equation}
\label{eqn:vertexcorrect}
\frac{\Gamma _{1}(i\omega ,ix)}{\Gamma _{0}}=\frac{U}{2\pi }\int ds\,
{\mathcal{G}}_{0}(i\omega -is){\mathcal{G}}_{0}(i\omega
-ix-is)D_{0}(is).
\end{equation}

\begin{figure}
\includegraphics[width=14cm]{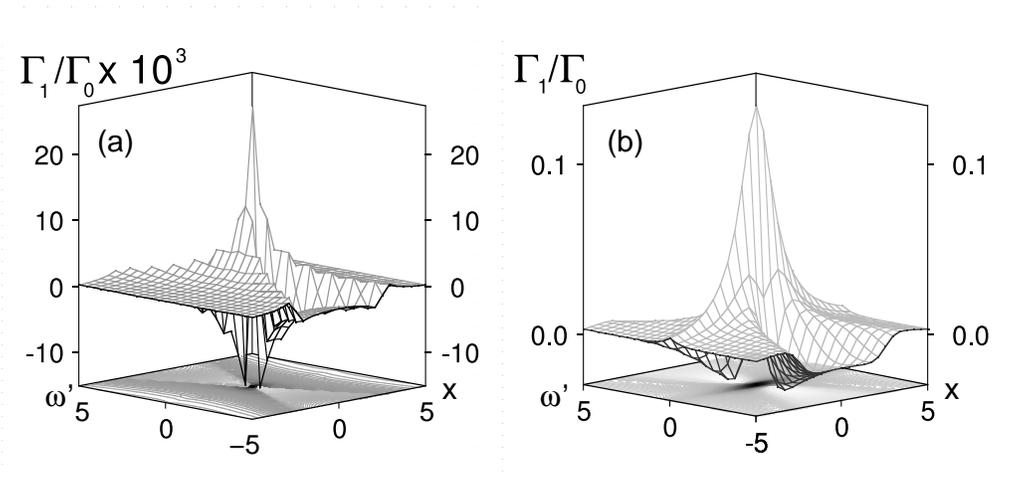}
\caption{The first order correction to the vertex function when (a)
$\omega_0 = 0.0556$ and $U = 0.214$ (b) $\omega _{0}= 0.75$ and
$U=0.214$. The central maximum in (b) shows a ratio of over $ 10\%$
and one should expect significant corrections to quantities computed
within ME theory. \label{fig:vertexcorrection}}
\end{figure}

In order for vertex corrections to be unimportant, the ratio $\Gamma
_{1}/\Gamma _{0}$ must be small. As an example, we show the lowest
order correction to the vertex function for (a) $\omega_0 = 0.0556$
and $U = 0.214$ (b) $\omega _{0}= 0.75$ and $U=0.214$ in figure
\ref{fig:vertexcorrection}. For the larger phonon frequencies, the
ratio is clearly greater than 10\% and the approximation is expected
to be significantly changed by the inclusion of vertex
corrections. For the weak coupling $U$ shown here, the corrections are
much less pronounced for small phonon frequencies.

\begin{figure}
\includegraphics[width = 10cm]{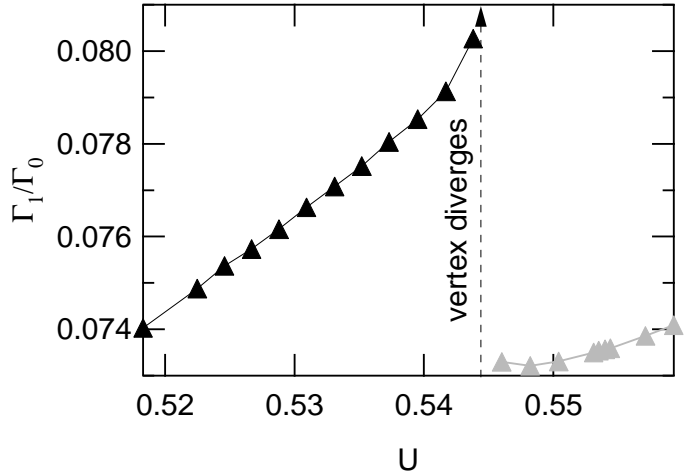}
\caption{Divergence in the vertex correction. At the
divergence Migdal's analysis breaks down. Points for larger values of
\protect\( U\protect \) (grey line) are shown in order to highlight
the divergence but the ME treatment used here is not valid above the
divergence, since the non-analyticity marks the limit of the expansion
in $\lambda$. \label{fig:diverge}}
\end{figure}

In Figure \ref{fig:diverge} we show the magnitude of the central peak
of the vertex correction (always the largest part of the function) for
a range of couplings at $\omega _{0}=0.0556$. Between $U=0.54$
and $U=0.55$ the vertex correction diverges, and ME theory clearly
breaks down. As it is necessary to pass through the divergence to
reach $U>0.545$, the theory is not applicable for strong
couplings.

\begin{figure}
\includegraphics[height=10cm,angle=270]{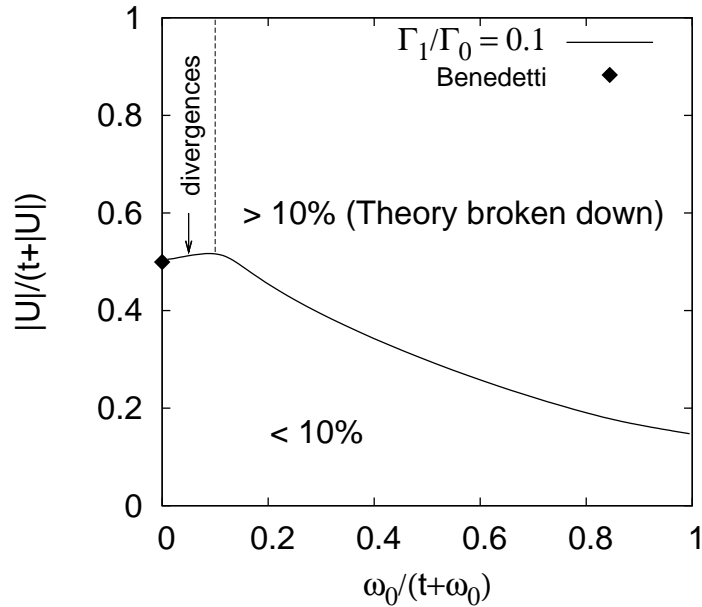}
\caption{Validity of ME theory as a function of $U'=U/(U+t)$ and
$\omega _{0}'=\omega_0/(\omega_0+t)$, chosen so that a value of 1
represents $U=\infty$, etc.  Also shown is the result of Benedetti and
Zeyher for very low phonon frequency (filled diamond). Above the line,
the ratio of the first order vertex correction to the bare vertex,
$\Gamma _{1}/\Gamma _{0}$, exceeds 10\% and ME theory is no longer
strictly valid. At low frequencies (adiabatic limit), breakdown occurs
at smaller coupling strength than expected as a result of the
divergence found in the vertex corrections. This is in contrast to the
interpretation of Migdal's analysis which is often used when low
frequency experimental data is analysed, and is the main result of
this article.
\label{fig:breakdown}}
\end{figure}

Figure \ref{fig:breakdown} shows the lowest value of \( U'=U/(t+U) \)
for given \( \omega _{0}' = \omega_0/(t+\omega_0) \) at which the
ratio \( \Gamma _{1}/\Gamma _{0} \) exceeds 0.1 (the primed quantities
are chosen so that $U'=1$ represents an unprimed value
$U=\infty$, and a primed value of $U'=0.5$ represents an unprimed value
$U=t$). Below this value ME theory can be expected to give accurate
results, while the theory begins to break down above this line. For \(
\omega _{0}'\gtrsim 0.2 \) ($\omega_0 \gtrsim 0.125$) the standard Migdal
criterion works well and correctly predicts the region in which ME
theory is applicable. However for frequencies \( \omega _{0}'\lesssim 0.2 \)
($\omega_0 \lesssim 0.125$), the Migdal criterion misses the divergence in
the vertex correction associated with the divergence in the effective
mass. This breakdown happens at $U\sim t$ or $\lambda\sim 1$. The line
of breakdown due to this vertex divergence levels off as the phonon
frequency approaches zero and does not scale as $1/U$ (Migdal's
criterion). These results imply that there is a breakdown in ME theory
in the adiabatic limit.

We note that at high phonon frequencies ($U,t\ll \omega _{0}$), our
line does not tend to zero. This is because the contribution $\Gamma
_{1}/\Gamma _{0}(x\rightarrow 0,\omega '\rightarrow 0)$ does not tend
to infinity. Instead, the function gets wider as the phonon frequency
increases (see figure \ref{fig:vertexcorrection}, panel b), and some
combination of the magnitude and width of the vertex is probably a
better measure of the importance of vertex corrections. At large
$\omega_0$, the breakdown is quite different to that at low phonon
frequencies as the second order diagrams (one of which contains a
vertex correction) develop the same functional form and
magnitude. However the breakdown at $U = 0.1$ ($U'=0.17$) is not too
bad an estimate.  At this value of $U$, there is only a small mass
renormalisation due to vertex corrections.

Our results are consistent with that of Benedetti and Zeyher
\cite{benedetti1}, which was obtained with a different technique using
a semi-circular density of states (${\mathcal{D}}_{SC}(\epsilon) =
\sqrt{4t^2 - \epsilon^2}/2\pi t^2$), and predicted breakdown for
$\lambda_{SC} \ge 1.25$ in the extreme adiabatic limit. We have
included their result in the figure, choosing the bandwidth parameter
by matching the bare DOS at the Fermi energy for the two cases (the
conversion factor is $\lambda = {\mathcal{D}}_{SC}(0) \lambda_{SC} /
{\mathcal{D}}(0) = \lambda_{SC} \sqrt{2/\pi}$) which is reasonable
when considering low energy excitations at half-filling, and
corresponds to a critical $\lambda \ge 0.997$ for the Gaussian DOS
used here (see the diamond in figure \ref{fig:breakdown}). Millis {\it
et al.}  \cite{millis2} also predicted a breakdown in ME theory at
$U\sim t$ from their solution of the static limit of the
infinite-dimensional Holstein model, consistent with the breakdown
that we have found. The solution of the static limit predicts the
formation of sub-bands, which is something not reproduced in the ME
solution. The lack of subbands in ME theory demonstrates that higher
order diagrams are essential for the description of the sub-band
formation, and therefore for the description of the strong coupling
limit.

We suggest that the breakdown of ME theory should be understood in the
following heuristic manner. As the coupling increases, the host
spectral function (\( {\mathcal{G}}_{0} \)) narrows and spectral
weight is shifted towards lower energy scales. There is then an
effective bandwidth, \( t_{\mathrm{eff}}<t \), for quasi-particles
close to the Fermi surface. The condition for the applicability of ME
theory then becomes $U\Omega\ll t^{2}_{\mathrm{eff}}\sim Z^2$. As seen
in figure \ref{fig:phonspec}, there is also a reduction in the
effective phonon frequency with increasing interaction
strength. Although the renormalisation of the phonon frequency helps
to drive against the transition, we note that the band narrowing
effect is much stronger, as can be seen by comparing figures
\ref{fig:qpweight} and \ref{fig:phonfreq}.

\section{Resistivity Curves}

\label{sct:rescurve}

\begin{figure}
\includegraphics[height = 10cm,angle = 270]{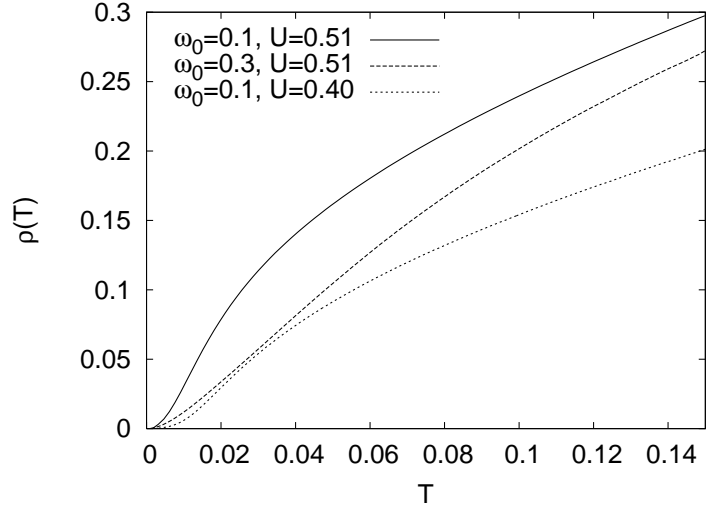}
\caption{Resistivity curves for various couplings and
temperatures. The resistivity is in units of $e^2V/ha^2$ with $V$ the
unit cell volume and $a$ the lattice cell spacing. The temperature at
which the shoulder appears is defined by the phonon frequency, and the
amplitude is defined by the coupling.  Low temperature behaviour is
\protect\( T^{2}\protect \) according to a weakly renormalized
electron gas. High temperature behaviour is linear. Note the negative
curvature at intermediate temperature. \label{fig:resistivityudep}}
\end{figure}

Using the finite temperature form of the self-energy, it is possible
to calculate resistivity curves. These are shown in figure
\ref{fig:resistivityudep} at various electron-phonon coupling and
phonon frequency. At high temperatures, the tendency is to linear
behavior, with a gradient depending on $U$, but independent of $\omega
_{0}$. The effect of phonon frequency is most dramatically seen at
$T\sim \omega _{0}/2$, where a point of inflection is seen, before the
curve tends to low temperature \( T^{2} \) behavior, consistent with a
weakly renormalized electron gas. We note that long range order has
not been considered in this study, so no superconducting transition is
seen. The negative curvature of the resistivity curve at intermediate
temperatures is of interest, since that curvature has recently been
interpreted as an onset of the Mott limit, but may have a different
interpretation as an intermediate coupling phenomenon \cite{lortz1}.

\section{Conclusions}

\label{sct:conclusions}

We have computed spectral functions, resistivity curves,
self-energies, quasi-particle weight and effective phonon frequency
for the Holstein model at half-filling using Migdal--Eliashberg
theory, within the DMFT framework, and revisited Migdal's analysis of
the strength of the vertex correction. By analysing the first order
vertex correction, we have defined the region of applicability of ME
theory. We find that ME theory breaks down at intermediate coupling in
the adiabatic limit, showing that Migdal--Eliashberg theory should
only be trusted at weak coupling within this framework. The coupling
at which breakdown occurs corresponds to a divergence in the effective
mass, indicating that both long range order and vertex corrections
should be included to correctly describe the strong coupling regime.

The question remains: What was lacking from the analysis of Migdal,
which estimated the magnitude of the vertex corrections. This can be
understood from the magnitude of the kinetic energy. In the
calculations, the kinetic energy can be seen to be decreasing due to
the localisation at the non-analyticities. Thus in Migdal's estimate,
the renormalised kinetic energy, and not the bare kinetic energy
should have featured. The renormalised phonon frequency should also
have been used, but it tends to zero more slowly than the kinetic
energy. At the non-analyticity, where the kinetic energy tends to zero,
it can be seen that a vertex divergence is expected from the modified
version of Migdal's estimate.

Our results are timely because of a recent upsurge of interest in the
role of electron-phonon coupling in the cuprate superconductors. There
are many experimental results which have been analysed and interpreted
using ES/ME theory, in spite of the large coupling constant that has
been predicted from that analysis. Many groups are using Migdal's
analysis to justify their claims, without a careful consideration of
the self-consistent effect of coupling on the kinetic energy of the
electrons (polarons). Exact numerical analysis shows that large
electron-phonon couplings significantly reduce the kinetic energy of
electrons \cite{hague2006b} and thus affect the internal consistency
of Migdal's analysis. Therefore, the any very large $\lambda$ determined
from ES style analysis of experimental results is not consistent with
a theory that neglects vertex corrections (i.e. a more detailed
analysis with higher order effects included is necessary). Without
that analysis, the only conclusion that can be reached is that the
coupling is large, but no reliable value for that coupling can be
determined. The main conclusion of a breakdown is not expected to
change in 1D, 2D or 3D. In three dimensional systems, the role of
spatial fluctuations should only be significant very close to the
bi-polaron instability. The role of spatial fluctuations on the
spectral properties in 2D has been analysed at weak coupling, and is
found to lead to quantitative differences in the coherent excitations
\cite{hague2003a}. Spatial fluctuations have a major role in the
superconducting state, and it has also been shown that the Eliashberg
approach to superconductivity is inadequate for optical phonon
mediated $d$-wave superconductivity \cite{hague2006a}, and incomplete
for $s$-wave superconductivity \cite{hague2005a}. We therefore urge
researchers to carefully analyse the internal consistency of their
theories when dealing with strong coupling materials.

\section{Acknowledgements}

The authors would like to thank F.Gebhard and F.Essler for useful
discussions. JPH thanks the EPSRC for partial funding of this work.


\pagebreak

\pagebreak

\pagebreak

\pagebreak

\pagebreak

\pagebreak

\pagebreak

\pagebreak

\pagebreak

\pagebreak

\end{document}